\pdfoutput=1

\documentclass[sigconf]{acmart}

\RequirePackage[l2tabu, orthodox]{nag}

\usepackage{dirtytalk}
\usepackage{lipsum}
\usepackage{siunitx}
\usepackage{listings}
\usepackage{ellipsis}
\usepackage[inline]{enumitem}
\usepackage{cleveref}

\usepackage[T1]{fontenc}
\usepackage[latin9]{inputenc}

\newcommand{\mem}{memory}
\newcommand{\oom}{\textsc{OOM}}
\newcommand{\oomk}{\textsc{OOM} killer}
\newcommand{\os}{\textsc{OS}}

\acmConference[ICSE 2024]{46th International Conference on Software Engineering}{April 2024}{Lisbon, Portugal}

\title{What Do You Mean by Memory?\texorpdfstring{\\}{}When Engineers Are Lost in the Maze of Complexity}

\begin{abstract}
An accepted practice to decrease applications' memory usage is to reduce the amount and frequency of memory allocations.
Factors such as
\begin{enumerate*}[label=(\alph*),before=\unskip{ }, itemjoin={{, }}, itemjoin*={{, and }}]
    \item
    the prevalence of out-of-memory (\oom) killers
    \item
    memory allocations in modern programming languages done implicitly
    \item
    overcommitting being a default strategy in the Linux kernel
    \item
    the rise in complexity and terminology related to \mem\ management makes the existing guidance inefficient.
\end{enumerate*}
The industry needs detailed guidelines for optimizing memory usage targeting specific operating systems (\os) and programming language types.
\end{abstract}

\begin{document}

\author[1]{Gunnar Kudrjavets}
\orcid{0000-0003-3730-4692}
\affiliation[obeypunctuation=true]{
   \institution{University of Groningen\\}
   \city{Groningen, }
   \postcode{9712 CP}
   \country{Netherlands}}
\email{g.kudrjavets@rug.nl}

\author[2]{Aditya Kumar}
\orcid{0000-0001-6312-2898}
\affiliation{
    \institution{Google}
    \streetaddress{1600 Amphitheatre Parkway}
    \city{Mountain View}
    \state{CA}
    \country{USA}
    \postcode{94043}}
\email{appujee@google.com}

\author[3]{Jeff Thomas}
\orcid{0000-0002-8026-9637}
\affiliation{
    \institution{Meta Platforms, Inc.}
    \streetaddress{1 Hacker Way}
    \city{Menlo Park}
    \state{CA}
    \country{USA}
    \postcode{94025}}
\email{jeffdthomas@meta.com}

\author[4]{Ayushi Rastogi}
\orcid{0000-0002-0939-6887}
\affiliation[obeypunctuation=true]{
   \institution{University of Groningen\\}
   \city{Groningen, }
   \postcode{9712 CP}
   \country{Netherlands}}
\email{a.rastogi@rug.nl}



\maketitle

\section{Introduction}

\paragraph{Primary motivation}

The accepted practice in software engineering is that every application should minimize its \mem\ usage to optimize performance~\cite{gregg_systems_2020,noble_2002}.
Optimal \mem\ usage helps an application achieve higher throughput, reduce the amount of paging or swapping for the \os, and decrease the resource requirements for the host system.
A critical reason in the modern \os s that causes applications to measure and control their \mem\ usage is the presence of an \oomk.
The \oomk\ is a system component that terminates applications when they \say{use too much \mem.}
\emph{When engineers improve \mem\ usage, they face challenges because the optimization techniques and terminology highly differ between \os s and types of programming languages}.
What the term \emph{\mem} means is highly context-dependent and requires a non-trivial amount of deciphering to interpret correctly.

\paragraph{Influx in complexity and terminology}

The evolution of hardware and \mem\ management techniques has caused an increase in the ways that \os s account for and classify \mem\ usage.
A classical textbook that describes the initial architecture of the \textsc{UNIX} spends \num{38} pages on \mem\ management~\cite{bach_1986}.
A modern overview of Microsoft Windows internals covers the same topic in \num{182} pages~\cite{yosifovich_2017}.
The author of a book on a contemporary overview of Linux kernel \mem\ management subsystem states as of writing this paper that \say{I have written \num{835} pages of a target of roughly \num{1200}--\num{1500} pages}~\cite{linux_mem}.
Though this is an anecdotal example, it corresponds with our industry experience while working on problems related to \mem\ management.
Thinking only at the scope of \texttt{malloc()} and \texttt{free()} has become obsolete and oversimplified.

We illustrate the increase in complexity by looking at the most famous ecosystems, such as commercial \os s that Apple and Microsoft produce and the open-source Linux kernel.
In~\Cref{tab:types-of-memory}, we enumerate \num{20} ways macOS and its derivatives, such as iOS, iPadOS, and watchOS, categorize \mem\ usage~\cite{singh_mac_2016,levin_ios_2017}.

\begin{table}[htbp]
    \caption{Different \mem\ usage quantifiers in Apple \os s.}
    \label{tab:types-of-memory}
    \centering

    \begin{tabular}{llll}
        \toprule
Anonymous & Mapped & Physical & Shared \\
& & footprint & \\
Clean & Memory & Private & Virtual \\
& footprint & &  \\
Compressed & Non-volatile & Purgeable & Volatile \\
Dirty & Persistent Bytes & Real & Wired \\
Evictable & Physical & Resident & Working set \\
        \bottomrule
    \end{tabular}
\end{table}

Microsoft Windows introduces even more definitions, such as commit size, paged pool, non-paged pool, and reserved \mem~\cite{yosifovich_2017}.
If that is insufficient to confuse an average software engineer, then Linux adds terms such as \textsc{PSS} (Proportional Set Size),
\textsc{RSS} (Resident Set Size),
\textsc{USS} (Unique Set Size), and
\textsc{VSZ} (Virtual Memory Size) to the mix~\cite{love_2013,love_linux_2005}.

\section{Industry challenges}

\begin{flushright}
\emph{\say{He who controls the amount of dirty pages in kernel controls the application's lifetime and its commercial success.}}
\end{flushright}
\begin{flushright}
--- A paraphrase of Frank Herbert's \say{Dune}~\cite{herbert_dune_1984}.
\end{flushright}

\paragraph{Scarcity of in-depth knowledge}

Most software engineers specialize in something other than \mem\ management internals for a particular \os.
Even fewer engineers have the in-depth knowledge that spans multiple \os s.
Nevertheless, most popular software systems like browsers, editors, and messengers support multiple mobile and desktop \os s.
A talk by Mark Russinovich, a \textsc{CTO} of Microsoft Azure, is appropriately titled \say{Mysteries of Memory Management Revealed}~\cite{russinovich_mysteries_2011}.
An operating systems engineer armed with a kernel debugger has become a modern-day mystic and sorcerer.

\paragraph{Invalidation of existing assumptions}

Starting from the initial versions of the \textsc{UNIX} in the early 1970s, the succinct guidance to optimize applications' \mem\ usage has been \say{call \texttt{malloc()} less.}
This recommendation has become obsolete with \os\ development advancements, increased complexity of \mem-related terminology, and heterogeneity in popular programming languages.
We find two primary paradigm shifts that invalidate the existing assumptions.
Firstly, \emph{overcommitting} in the popular Linux kernel means that \os\ only allocates \mem\ when the allocated pages are written to by a consumer~\cite{love_2013,love_linux_2005}.
As a result, the frequency and size of allocations have lost their original meaning.
Secondly, high-level programming languages have abstracted \mem\ management away from the programmer.
Techniques such as garbage collection have made allocations and deallocations \emph{implicit}, seamless, and non-deterministic.

\paragraph{Need for mapping between intents and actions}

Different platforms share the standard performance engineering goals related to \mem.
Engineers want to
\begin{enumerate*}[label=(\alph*),before=\unskip{ }, itemjoin={{, }}, itemjoin*={{, and }}]
    \item
    avoid premature termination of their applications
    \item
    stay under a certain quota or limit of \mem\ usage
    \item
    optimize the metrics that matter in a particular context.
\end{enumerate*}
Unfortunately, the current published guidance is either overtly general or limited.
\emph{We observe repeated rediscovery of the same facts and that critical knowledge has become limited to a tiny group of engineers}.
\Cref{apple-and-ios} provides one concrete example showing the ambiguity and complexity of daily engineering tasks for iOS.

\section{A case of Apple and iOS}
\label{apple-and-ios}

We will use iOS as a concrete example due to the popularity of Apple's ecosystem.
As practicing engineers, we have witnessed similar challenges with Android, various Linux distributions, and Microsoft Windows.
In February 2023, Apple stated, \say{\dots\ we now have more than 2 billion active devices as part of our growing installed base}~\cite{apple_device_count}.
A significant portion of these devices are cell phones that use iOS.
Most modern \os s support paging (writing \mem\ pages that \os\ does not actively use into a secondary storage such as disk)~\cite{anderson_operating_2014}.
iOS does not support paging out~\cite{levin_ios_2017,apple_paging}. %

As a result, \mem\ is one of the most precious system resources on iOS.
The \oomk\ will terminate the iOS applications that exceed a specific limit.
Therefore, each iOS application developer must understand how to prevent a situation where an \oomk\ terminates the application.

An engineer who wants to understand what specific criteria \oomk\ uses to terminate applications must read the \textsc{XNU} kernel source code.
After the discovery process, the engineer will hopefully reach the definition in~\Cref{code:phys_footprint} that describes the application's physical footprint~\cite{xnu_kernel}.
Like a Russian Matryoshka doll, the engineer now faces a new set of conundrums to unwrap.

\lstset{language=c,breaklines=true,frame=single,basicstyle=\ttfamily\footnotesize,caption={Definition of physical footprint in \textsc{XNU} kernel.},label=code:phys_footprint}
\begin{lstlisting}
Physical footprint: This is the sum of:

+ (internal - alternate_accounting)
+ (internal_compressed - alternate_accounting_compressed)
+ iokit_mapped
+ purgeable_nonvolatile
+ purgeable_nonvolatile_compressed
+ page_table
\end{lstlisting}

How do {Objective-C} or {Swift} developers apply these findings?
Both languages, by default, use Automatic Reference Counting (\textsc{ARC}).
\textsc{ARC}, by design, abstracts \mem\ management away from engineers.
What does \emph{reducing the amount of purgeable non-volatile compressed \mem} they consume mean?
What are the \say{good} and \say{bad} amounts?
Should the engineers create fewer objects, write less code, do less of everything, or more of something else?
How to even apply any typical advice to optimize \mem\ usage in programming languages like this~\cite{gregg_systems_2020}?

\section{Industry needs}

The complexity of \mem\ management means long-term job security for operating systems and performance engineers.
However, the current situation is detrimental to most software engineers who do not specialize in those fields.
Determining what subset of metrics matters and how to control them to achieve a desired outcome has become a costly and involved effort.
Non-specialists need \emph{concrete guidance that helps them to map their intents to specific actions depending on the context}.
With the rise in popularity for high-level languages such as Java, Kotlin, Python, or Swift where the \mem\ management is \emph{not explicit}, the lack of guidance has become problematic.

In our industry experience, the set of typical questions that the engineers who face \mem-related performance challenges ask are the following:

\begin{enumerate}
    \item
    What do any of these metrics mean?

    \item
    Which metrics are essential in the context of a particular application, programming language, and \os?

    \item
    What is the desired range for a specific metric for a given application?

    \item
    How do I modify the application to change the critical metrics in the desired direction?
\end{enumerate}

The research community can help practitioners by
gathering experimental data to
\begin{enumerate*}[label=(\alph*),before=\unskip{ }, itemjoin={{, }}, itemjoin*={{, and }}]
    \item
    determine what metrics matter (or not) for a particular intent on a specific \os

    \item
    how to evaluate certain metrics for programming languages that manage memory implicitly

    \item
    provide guidance for languages that use garbage collection
\end{enumerate*}

Another vital contribution is the case studies describing how practitioners have solved specific problems.
A small number of software companies are developing the majority of popular applications.
Sharing the existing in-house knowledge will help advance the software performance engineering field.

\bibliographystyle{ACM-Reference-Format}
\bibliography{dystopian-memory}

\end{document}